# Spring Festival points the way to cleaner air in China

Siwen Wang[1], Hang Su[1]*, David G. Streets[2], Qiang Zhang[3], Zifeng Lu[2], Kebin He[4], Meinrat O. Andreae[1,5], Ulrich Pöschl[1] and Yafang Cheng[1]*

[1]Max Planck Institute for Chemistry, Mainz 55128, Germany

[2]Energy Systems Division, Argonne National Laboratory, Lemont, IL 60439, USA

[3]Ministry of Education Key Laboratory for Earth System Modeling, Department of Earth System Science, Tsinghua University, Beijing 100084, China

[4]State Key Joint Laboratory of Environment Simulation and Pollution Control, School of Environment, Tsinghua University, Beijing 100084, China

[5]Scripps Institution of Oceanography, University of California San Diego, La Jolla, CA 92093, USA

*e-mail: yafang.cheng@mpic.de; h.su@mpic.de

Human migration during the Chinese Spring Festival (SF) is the largest collective human activity of its kind in the modern era—involving about one-tenth of the world population and over six percent of the earth's land surface area. The festival results in a drop of air pollutant emissions that causes dramatic changes of atmospheric composition over China's most polluted regions. Based on satellite and *in-situ* measurements for the years 2005–2019 over 50 cities in eastern China, we find that the atmospheric $NO_2$ pollution dropped by ~40% during the SF week, and fine particulate matter ($PM_{2.5}$) decreased by ~30% in the following week, reflecting the effectiveness of precursor emission controls on the mitigation of secondary $PM_{2.5}$ formation. However, although human activity and emissions are at the lowest level, air pollution over eastern China during the SF still far exceeds that over other worldwide pollution hotspots. Our analyses suggest that measures based solely on end-of-pipe controls and industry upgrades may not suffice to meet air quality goals. Further cleaning of the air in China depends fundamentally on sustainable advances in both heavy industry upgrades and clean energy transition.

Rapid urbanization has been one of the primary driving forces of China's economic miracle during the past two decades[1,2]. It has greatly improved the living quality of the majority of the population, but has led to severe air pollution issues[3–5] with high epidemiological risks[6–8] and jeopardized the sustainability of economic development[9,10]. Improving air quality has been an important motivation for China's transition in industrial and energy structures in recent years to avoid irreversible losses of public health and climate benefits. Making effective mid- and long-term policy decisions on air pollution controls rests on three major branches of scientific enquiry: (i) the sources and rates of primary air pollutant emissions and the reduction potential through the phase-out of outdated facilities and the introduction of alternative technologies and clean energy; (ii) the characteristics of local atmospheric dynamics and chemistry (evolution and sinks) acting on ambient air pollutants and the responses of secondary species (e.g., $PM_{2.5}$ and Ozone ($O_3$), which have higher health risks) to different control measures and policy emphases; and (iii) the marginal treatment costs and complexity of different measures in implementation and management.

Acquiring knowledge on these issues requires systematic field experiments involving dynamic source investigations, intensive atmospheric composition monitoring, and effective measures causing significant emission changes. Such opportunities are scarce and costly, but may be occasionally realized during high-ranking international and national events when temporary mandatory emission controls (TMECs) are implemented, including halting local and nearby industrial and construction activities and vehicle restriction. Typical examples are the 2006 Sino-African Summit[11,12] and the 2008 Beijing Olympics[13,14]. Here, we show that in contrast to the irreproducible and costly TMEC events, the annual Spring Festival (SF) provides a unique time window (as a natural field laboratory) to examine the characteristics of emission sources and the atmospheric composition responses to the rapid emission changes in urban areas of different socioeconomic and meteorological conditions that related to reproducible short-term human migration activities throughout China.

The SF period includes the world's largest human long-distance migration activities involving about 3 billion trips within a 40-day period (Supplementary Information). During this important traditional festival, the returning of the Chinese to their hometowns from their normal residences (a reversal of urbanization) results in a decrease of high-emitting light industrial activities (e.g., brick, ceramic, lime, and textile industries), as workers flow away, and the disappearance of traffic jams in the metropolises. Atmospheric observations show a significant reduction of air pollution during the SF period over China corresponding to a reachable emission reduction level that could be used as a reference to the future cleaner air in China.

In this study, daily Aura-Ozone Monitoring Instrument (OMI) tropospheric $NO_2$ columns and total $O_3$ columns for 2005–2019 from the National Aeronautics and Space Administration (NASA), and the *in-situ* $PM_{2.5}$, $NO_2$, and $SO_2$ data for 2013–2019 from the Ministry of Ecology and Environment (MEE) of China monitoring networks have been analyzed. Nitrogen oxides ($NO_x = NO + NO_2$) are emitted by all combustion processes and have been demonstrated to be an ideal indicator for anthropogenic emissions by satellite instruments[3,15–17]. Due to the irregular dates of the SF on the Gregorian calendar, we define the SF day as the origin (day 0) of each lunar year, and the daily averages of data are aligned according to their lunar calendar dates for the studied years (Supplementary Information). This temporally aligns the intrinsic shifts of human activities before and after the SF in different years. In contrast to previous studies that focused on a single year or a specific city[18–23], random interferences (e.g., retrieval biases of satellite and *in-situ* measurements and meteorological variabilities) are greatly neutralized in the multi-year averages with temporal alignment, and thus the systematic changes in short-term air pollution emissions and non-emission-related factors could be characterized.

Figure 1a shows the reduction ratios in the OMI tropospheric $NO_2$ columns of the SF week (days 0–6) over parts of East Asia at a resolution of $0.25° \times 0.25°$, compared to the column averages of the third week before the SF (see Methods and weekly evolution of $NO_2$ pollution in Extended Data Fig. 1). The majority of significant reduction is observed over vast areas in eastern China, in particular over the "GDP-Top-50" cities (Extended Data Table 1) located in the North China Plain (NCP), Yangtze River Delta (YRD), Pearl River Delta (PRD), central and northeastern China, and the Sichuan Basin. Similar reductions are also seen over South Korea, which celebrates the SF as well. However, the daily evolution of tropospheric $NO_2$ columns exhibits significant differences among grid categories (Fig. 1b), with a distinct "$NO_2$ hole" over the developed city clusters (>40% decreases) around the SF week, moderate decreases over typical heavy industrial and resource-driven cities[24] (~10–30%), and small changes (<10%) over isolated power plants in rural areas[25] (Extended Data Table 2). This reflects the differentiation in the sectoral emission changes and the local dominant sources, and can be used to infer the emission source structure and the reduction potential (Supplementary Information).

To further elucidate the origins of these systematic changes during the SF and examine the responses of secondary species (e.g., $O_3$ and $PM_{2.5}$) to primary emission reductions, the daily evolutions of OMI tropospheric $NO_2$ columns and total $O_3$ columns were normalized for the "GDP-Top-50" cities in Fig. 2a. The baseline variations of the two indexes (dashed lines) due to non-emission-related factors are generated from multi-year chemical transport model (CTM) simulations with fixed emission rates of air pollutants, reflecting the systematic trends in $NO_x$ lifetime (declining due to enhanced photochemical losses) and seasonal $O_3$ cycles (springtime maximum in northern mid-latitudes due to stratospheric meridional circulation[26] and tropospheric photochemical production[27]). The tropospheric $NO_2$ columns drop rapidly following the beginning of the human migration (~10 days before the SF day), reach a minimum in the SF week, and rebound slowly to the baseline level after about four weeks. A small decline of $NO_2$ is also notable around the Lantern Festival (LF, day 14). This specific pattern was found to be highly correlated (correlation coefficient ($R$) of 0.93 for days –7 to 21) with the congestion delay index during that period (Extended Data Fig. 2a), an index reflecting the remaining urban transportation (or rather the human activities) estimated based on the traffic data for Chinese cities[28]. Statistical activity data suggest an average 21% (±2%) reduction in the national industrial electricity consumption for the SF months during 2010–2019 and a 9% (±6%) increase in the residential component[29] (Extended Data Table 3). These pieces of evidence reveal the dominant role of the human migration-related emission changes in the reduced $NO_2$ signals over urban areas during the SF.

Overall, the average net reduction of the tropospheric $NO_2$ columns is estimated to be 22% during the whole SF period (shaded area in Fig. 2a), peaking at 40% in the SF week, comparable to the statistical activity reductions mentioned above. The total $O_3$ columns show a similar evolution as the tropospheric $NO_2$ but with a lag period of 1–2 weeks due to the longer lifetime of tropospheric $O_3$ (see Fig. 2b). The average reduction of the total $O_3$ column is 1.7% during the SF, corresponding to about 5 Dobson Units (DU) or 14% of the tropospheric component, and peaks at 2.7% (8 DU or 22% of the tropospheric component) in the week following the SF week (tropospheric $O_3$ calculated according to the modeled vertical profiles, see Methods). The effectiveness of precursor emission reductions for containing the winter $O_3$ pollution is comparable to that during the 2008 Beijing Olympics[14].

The *in-situ* measurements also show reductions in surface $NO_2$ concentrations (44%) comparable to the column values for the "GDP-Top-50" cities during the SF week (Extended Data Fig. 2b). Meanwhile, surface $SO_2$ decreased by 32%. Correspondingly, the surface $PM_{2.5}$ declined by ~30% a week later (temporary $PM_{2.5}$ increases around the SF day reflecting the fireworks activities[21–23]), showing a lag period due to the longer lifetime of aerosols. The distinct lag period of $PM_{2.5}$ (and $O_3$ mentioned above) implies that the implementation of TMECs for high-ranking events and the early warning mechanisms of air pollution in Chinese cities may have limited effects on the prompt reduction of secondary aerosol formation, which can account for ~30–77% of ambient $PM_{2.5}$ during haze events there[30]. Further, a strong linear relationship (*R* of 0.97) was found between the descending periods of surface $PM_{2.5}$ and the sums of $NO_2$ and $SO_2$ during the SF over cities in the NCP, YRD, and PRD regions (Fig. 2c). Our long-term statistical results provide direct observational evidence that emission reductions in primary air pollutants can have near-linear mitigation effects on wintertime $PM_{2.5}$ pollution over the most polluted areas in China.

In addition to proving the effectiveness of large-scale pollution "controls", the Chinese Spring Festival could also give us great inspiration in terms of pollution control potential and future clean air policies in China. As shown in Fig. 3a–c, even though urban human activities and air pollution over eastern China are at their lowest levels in wintertime, the average $NO_2$ levels in China during the SF weeks in 2005–2019 were still much higher compared to population and pollution hotspots in the eastern US and western Europe. Distributions of the gridded OMI tropospheric $NO_2$ columns for the selected domains in China suggest dense $NO_2$ pollution over the NCP (twice that of the other regions). The OMI planetary boundary layer (PBL) $SO_2$ columns and *in-situ* surface $PM_{2.5}$ observations during the SF of 2013–2019 also show the remaining differences in air pollution controls between the NCP and developed western countries (Extended Data Fig. 3, i.e., 3–6 times higher $PM_{2.5}$ over the NCP). It implies that even the massive reduction in anthropogenic air pollutant emissions (up to ~40% during the SF week) that was attained by minimizing urban industrial and vehicle activities is still quite far away from meeting air quality goals in China. Another extreme example for the suppression of human activities and the corresponding air pollution is the SF in 2020 which coincided with the outbreak of COVID-19 throughout China, as reflected by the even more significant drop in $NO_2$ pollution (Extended Data Fig. 4), however, with destructive economic damages.

Cleaner air has always been one of the major drivers for emission controls in China. In retrospect, China has substantively promoted technological improvements to upgrade its industrial structure (named *supply-side reforms*, with emphasis on increasing the shares of advanced manufacturing and high-technology industries and a synergistic reduction in industrial energy consumption per unit output; see the sectoral energy use in Fig. 4). Stringent air pollution control measures were taken to cut down the end-of-pipe emissions from coal-fired power plants, steel, petrochemical, and cement production (e.g., the installation of desulfurization, denitrification, and dust-removal devices) and vehicles, or to directly phase out small high emitters. In recent years, China has also activated energy structure reforms, with pledges to raise the shares of natural gas and non-fossil energy in the primary energy mix to 15% and 20% in 2030, respectively[31]. An important step was the implementation of actions to clean up residential heating (replacing coal use with natural gas or electricity) since 2017 in northern pilot cities[32]. As a result of these efforts, the national annual $PM_{2.5}$ in 2018 decreased by 48% compared to 2013 (from 105 to 54 µg m$^{-3}$), while the total energy consumption increased by 12% (Fig. 4). Meanwhile, the frequencies of mid- and high-polluting days[33] were significantly reduced. However, even after all these measures, the current $PM_{2.5}$ levels in China are still much higher than those in the US and Europe, or the air quality guideline recommended by the World Health Organization (WHO). During the SF period from 2013–2019 (Extended Data Fig. 5), a significant reduction of tropospheric $NO_2$ columns can be found, but the average column burden over the NCP is still twice as much as that in the similarly populated regions in the US and Europe.

Since human activities are reduced to their lowest level with up to ~40% total emission reductions, it is very likely that air pollution during the SF can represent the best possible results of emission reductions from regular end-of-pipe controls and industry upgrades. Modeling evaluation based on the existing energy transition plans and high penetration of advanced end-of-pipe control technologies confirmed that an ultimate >50% reduction in primary emissions cannot drop the annual mean $PM_{2.5}$ to a 35 μg m$^{-3}$ level in the Beijing-Tianjin-Hebei (BTH) regions by 2030 (ref.[34])—an air quality level still higher than that in the US and Europe. The remaining stubborn urban air pollution in eastern China during the SF is mainly attributable to coal-fired power plants and heavy industries, especially in heavy industrial and resource-driven regions (Fig. 1b). The reason for this mitigation shortfall lies in the high consumption of fossil fuels and the coal-dominated energy structure in China (see the doughnut charts in Fig. 4 for cross-regional comparisons[35]). It is estimated that about 70% of current coal-fired power plants in China have been already operated close to the designed standards of "ultralow" emissions[36]. In 2018, non-fossil energy already accounted for about 15% of Chinese primary energy, which leaves limited prospects for the development of non-fossil energy to the target share of 20% and the corresponding air quality benefits in the coming decade. With the implementation of stricter vehicle emission standards and more proactive incentives on fleet electrification (e.g., electric vehicle penetration at a level of 50%), urban vehicle emissions could be reduced by up to 40% by 2030 (though additional emissions from power generation would increase)[37], merely comparable to the reduction level of vehicle activities and emissions during the SF period. Therefore, to achieve the goal of clean air in China, in addition to further promoting end-of-pipe controls and industrial structure optimization, a pervasive energy structure reform is imperative.

In conclusion, our results show that emission mitigation measures based solely on end-of-pipe controls and industry upgrades may not suffice to meet China's air quality goals. Further cleaning of the air depends fundamentally on sustainable advances on both the heavy industry upgrades and energy structure reforms. Besides the promotion of vehicle fleet electrification, future measures should give more attention to the energy transition in power and heavy industry sectors to cut down the persistent high coal consumption. The total capacity of natural-gas-fired power generation has increased by a factor of six during 2013–2018 (from 11.6 to 83.8 gigawatts (GW)), but still accounts for only 3% of the annual electricity generation[27]. There is still great potential for substituting coal with natural gas in both heavy and light industries in China, although technical and market barriers must be overcome. The energy transition in the power and industry sectors needs to accelerate primarily in urban areas with more supportive fiscal and tax incentives and more obligations from the State-owned enterprises. Additional assessments of city-level emission source activities and development of next-generation (geostationary) satellite instruments can improve the reliability of analyses at the local level and at higher spatial resolution[16] and thus support more effective policy decisions on air pollution controls. Economic and management tools are also worth being introduced into the policy-making process concerning air pollution controls and measures, in order to achieve a more holistic optimization.

**METHODS**

**Satellite measurements.** Tropospheric $NO_2$ columns, total $O_3$ columns and PBL $SO_2$ columns from OMI are provided by NASA. The OMI onboard the NASA Earth Observing System (EOS) Aura spacecraft (http://aura.gsfc.nasa.gov/) was launched on 15 July 2004 into a near polar, sun-synchronous orbit at 705-km altitude, with a 13:45 local equator-crossing time[38]. OMI is a Dutch-Finnish nadir-viewing space-borne hyperspectral imaging spectrometer measuring the complete spectrum in the UV/VIS wavelength range (270–500 nm) with a high spatial resolution ($13 \times 24$ $km^2$ at nadir) and daily global coverage[38,39].

The operational Level 3 products (version 3) were used in this study. The OMI Level 3 data are produced by using the best pixel data over small equal angle grids ($0.25\degree \times 0.25\degree$) covering the whole globe (see refs.[40,41] for details of data processing and validation). The abundance of atmospheric $NO_2$ and $O_3$ are quantified along the viewing path (slant column) using a DOAS (Differential Optical Absorption Spectroscopy) algorithm[39,42,43], and converted to vertical columns by air mass factors (AMFs)[44] calculated from radiative transfer models. The fitting error in the $NO_2$ slant column is estimated to be less than $1 \times 10^{15}$ molecules $cm^{-2}$ (ref.[45]). The total $O_3$ columns show a 1.1 DU standard error compared with ground-based remote sensing measurements[46]. PBL $SO_2$ columns are derived using an improved Band Residual Difference Algorithm (BRD)[47] with local monthly AMFs from the GEOS-Chem model, and were used only for auxiliary analysis in this study due to poor data availability at a daily level. We have also examined our analyses with stricter data screening criteria[17] using the Level 2 orbit $NO_2$ data from NASA, and found similar results as presented here.

The daily column data were aligned according to their lunar calendar dates and averaged for each lunar day before and after the SF for 2005–2019. We chose the averages of column data of the third week before the SF (days –21 to –15) to normalize the pollution profile or calculate the reduction ratio for each $0.25\,°$ grid. The pollution of a city or an isolated power plant was represented by columns of a single grid where the city centers or point sources locate (coordinates provided in Extended Data Tables 1 and 2). The average column profiles of tropospheric $NO_2$ and total $O_3$ of the "GDP-Top-50" cities (Fig. 2a) were calculated based on the normalized column profiles of each city. The differences in city scales have not been considered in this study. Involving more grids (e.g., ~80 $\times$ 80 km$^2$ for each city) into analyses would dilute the distinct urban air pollution evolutions. The average adjustment factor for tropospheric $NO_2$ column reduction estimates during the SF week (Extended Data Table 1) was 0.92 as marked in Fig. 2a.

**Ground-based measurements.** The hourly *in-situ* measurement data for $PM_{2.5}$, $NO_2$, and $SO_2$ are recorded at more than 1,500 atmospheric monitoring sites managed by the MEE of China since 2013. These urban stations geographically cover all prefecture-level cities of mainland China. Each site reports the average concentrations of air pollutants over the surrounding 0.5–4.0 km (in radius) areas[48]. The daily 24-hour average concentration data were processed for each site using similar methods as of the satellite columns. We applied data from 200, 126, and 58 sites located within the NCP (35–41 °N, 113–120 °E), YRD (29–33 °N, 119–122 °E), and PRD (22–23.5 °N, 112–114.5 °E) regions, respectively. MEE measurements before 2013 have not been used due to significantly inconsistent sampling amounts and data quality (also with no daily $PM_{2.5}$ concentration data).

**Chemical transport model.** The GEOS-Chem model version 11-01 was applied to generate the baseline variations of air species during the SF period for 2005–2019. GEOS-Chem is a global chemical transport model for air composition studies (http://www.geos-chem.org/), driven by assimilated meteorological data from the Goddard Earth Observing System (GEOS) of the NASA/Global Modeling and Assimilation Office (GMAO), with 72 vertical hybrid sigma-pressure levels (14 layers in the lowest 2 km) extending up to 0.01 hPa. The model was performed for a global simulation at a 2.5 ° longitude × 2 ° latitude resolution with 47 vertical layers (in reduced mode), compiled with the "SOA" chemical mechanism to include the formation of secondary organic aerosols. Vertical mixing in the PBL follows a non-local scheme[49] that accounts for the varying magnitude of mixing from stable to unstable states of the boundary layer[50]. We used the MERRA-2 reanalysis meteorological data product (updated every 1–3 hours) in our simulation, which covers a continuous period from the year 1980 onwards. At least 1-year spin-up simulation was included to remove the effects from the initial concentration fields.

To evaluate the systematic effects from non-human-activity-related factors on the daily air pollutant evolutions, we conducted the GEOS-Chem simulations with fixed anthropogenic emissions over China for 2005–2019. The global anthropogenic emissions are provided by the Emission Database for Global Atmospheric Research (EDGAR version 4.2) inventory for 2012 (refs.[51,52]), and were replaced over China by the Multi-resolution Emission Inventory for China (MEIC) inventory (version 1.3)[53–55]. The gridded daily emission fluxes of $NO_x$, $SO_2$, $NH_3$, and primary $PM_{2.5}$ components for the first day of 2012 were used. The daily early afternoon (13:00–15:00 local time) tropospheric $NO_2$ columns and total $O_3$ columns were derived, normalized, and averaged over the "GDP-Top-50" cities to generate the baseline variations for OMI column changes during the SF period (Fig. 2a), which reflect the systematic trends in $NO_x$ lifetime and seasonal $O_3$ cycles[26,27]). Monthly tropopause heights were used to calculate the tropospheric columns. The daily 24-hour average concentrations of $PM_{2.5}$, $NO_2$, and $SO_2$ were derived, using a similar method, for the surface layer of the model, which typically corresponds to a height of about 120 m in winter over eastern China. Simulations with different emission rates (e.g., the MEIC emissions for the first day of 2016) produced no significant biases on the calculated baselines.

**Data availability.** OMI measurements are adopted from the National Aeronautics and Space Administration's (NASA) Goddard Earth Sciences (GES) Data and Information Services Center (DISC) (https://disc.gsfc.nasa.gov/). The hourly *in-situ* air quality data from the Ministry of Ecology and Environment (MEE) of China monitoring stations are accessible from https://beijingair.sinaapp.com/.

**Acknowledgements** This work is supported by the Max Planck Society (MPG). Y.C. acknowledges the Minerva Program of MPG.

**Author contributions** Y.C., H.S. and S.W. conceived the study. S.W. performed the research, including data analyses, model simulations and figures. S.W., Y.C. and H.S. analysed the results. D.G.S, Q.Z., Z.L, K.H., M.O.A. and U.P. commented on the manuscript. S.W., Y.C.

and H.S. wrote the paper with contributions from all co-authors.

**Competing interests** The authors declare no competing financial interests.

**Supplementary Information** is available for this paper.

**Materials & Correspondence** Correspondence and requests for materials should be addressed to Y.C. (yafang.cheng@mpic.de) or H.S. (h.su@mpic.de).

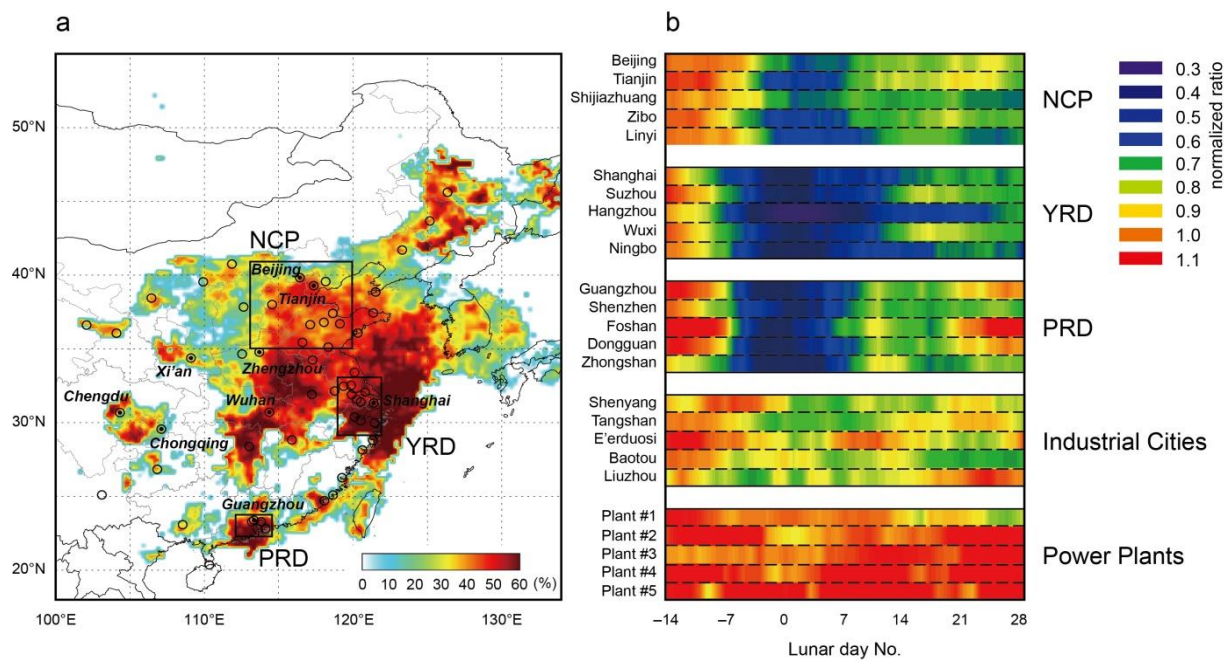

**Figure 1 | Mitigation of air pollution over eastern China during the Spring Festival (SF).**
**a,** Reduction ratios (%) in the average tropospheric $NO_2$ columns during the SF week compared to the column averages of the third week before the SF for the years 2005–2019 (Methods). The tropospheric $NO_2$ column data are observed by Aura-OMI at a resolution of $0.25° \times 0.25°$. Remote areas with grid average columns $< 2 \times 10^{15}$ molecules $cm^{-2}$ are excluded. Circles denote the "GDP-Top-50" cities (Extended Data Table 1) and provincial capital cities, with nine national central cities marked by black dots. Rectangles show the domains of three major city clusters, the North China Plain (NCP), the Yangtze River Delta (YRD), and the Pearl River Delta (PRD). **b,** Normalized daily evolution of tropospheric $NO_2$ columns over major cities from the three city clusters, typical industrial and resource-driven cities (classified according to ref.[24]), and isolated power plants[25] (Extended Data Table 2). Data are aligned according to the Chinese lunar calendar (see Supplementary Information) with a 7-day moving-average for each site.

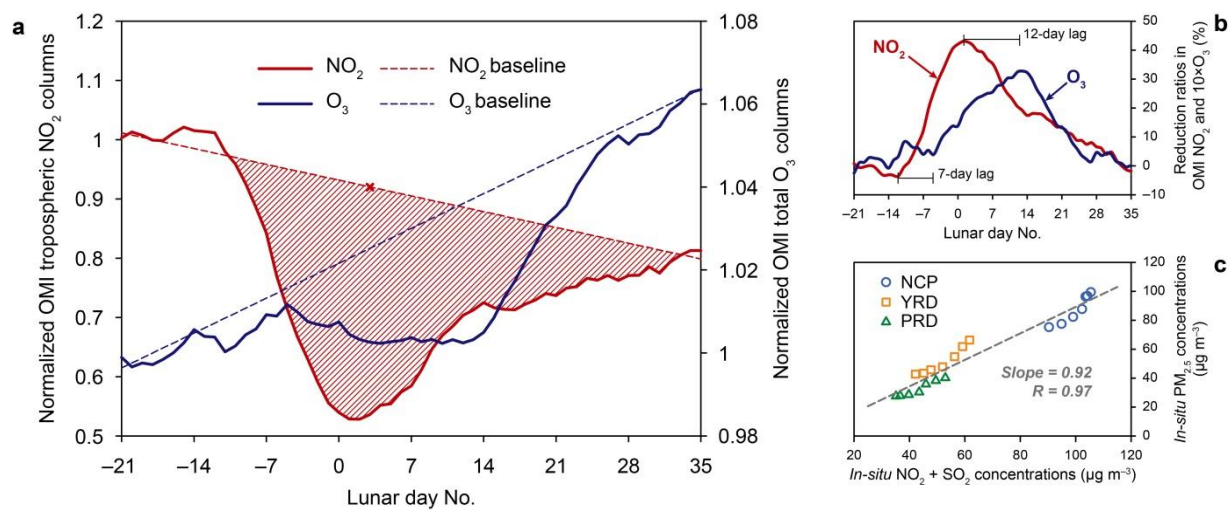

**Figure 2 | Pronounced decrease of urban air pollution driven by human migration during the SF period. a,** Average profiles of the normalized daily variations of Aura-OMI tropospheric $NO_2$ columns and total $O_3$ columns over the "GDP-Top-50" cities for 2005–2019. Dashed baselines are generated by the chemical transport model (CTM) with fixed emission rates of air pollutants (Methods). Shaded areas show the net reductions in tropospheric $NO_2$ columns during the whole SF period. Data are normalized and 7-day moving-smoothed for each city before averaging. Point (marked x) denotes the average adjustment factor for $NO_2$ reduction estimates during the SF week (Methods). **b,** Reduction ratios (%) in tropospheric $NO_2$ columns and total $O_3$ columns calculated from data in **a**. **c,** Response of fine particulate matter ($PM_{2.5}$) concentration to the reduction of its gas phase precursors. The scatterplot shows the *in-situ* $PM_{2.5}$ concentrations over city clusters in Fig. 1a versus that of the sums of $NO_2$ and $SO_2$. For the given regions, daily *in-situ* concentrations of species during the descending period before the SF (days –6 to 0 for $NO_2$ and $SO_2$, and 0 to 6 for $PM_{2.5}$ to consider a 1-week lag) are used for the analyses. The dashed line in **c** presents the linear regression fit using all samples.

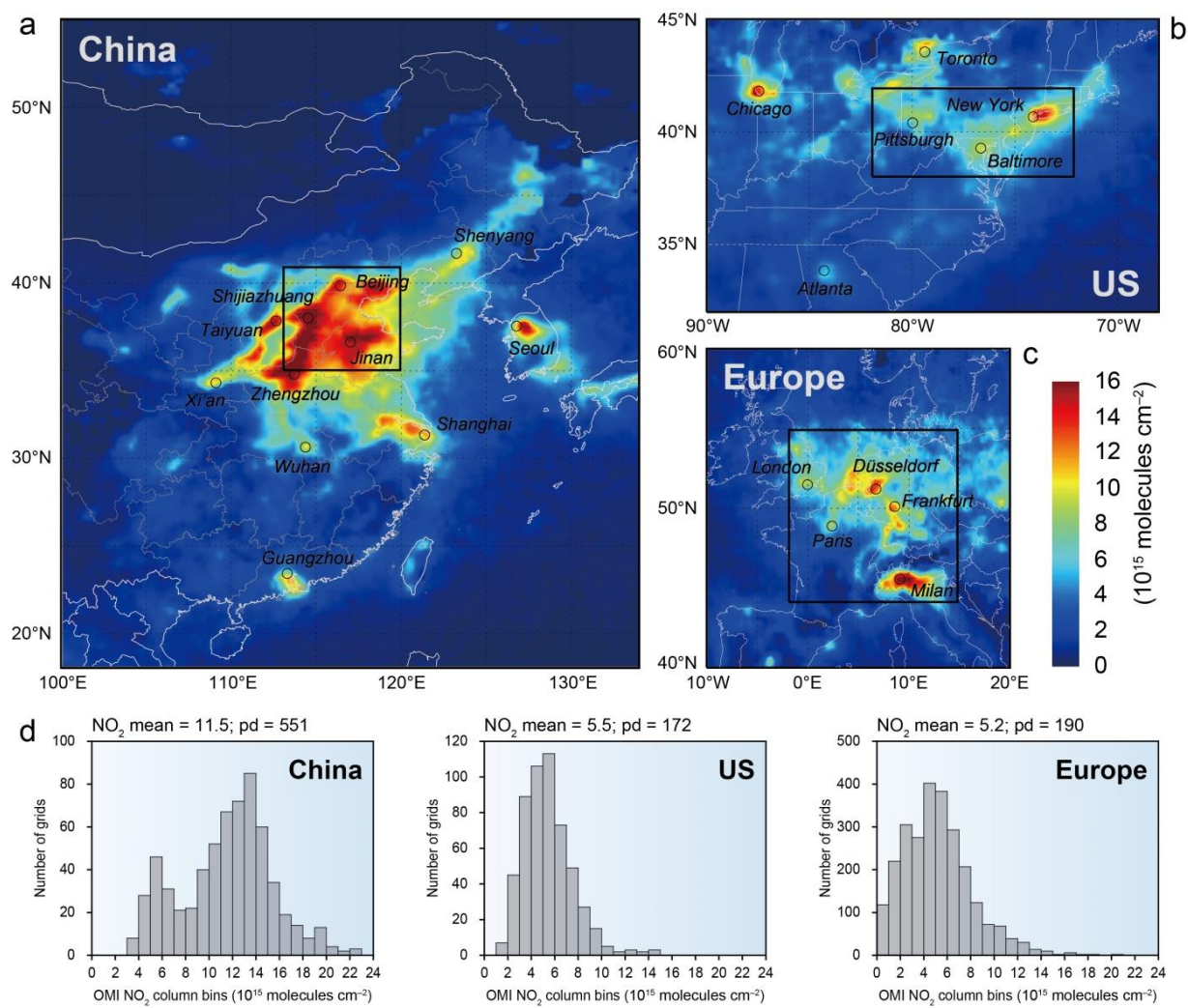

**Figure 3 | The NO$_2$ pollution during the SF was still high over eastern China. a–c,** Aura-OMI tropospheric NO$_2$ columns averaged over the SF weeks of 2005–2019 for three mid-latitude regions in the North Hemisphere (**a**, eastern China; **b**, eastern US; **c**, western Europe). Circles denote the major cities covered by the maps. **d,** Distribution histograms of the gridded OMI tropospheric NO$_2$ columns in the selected domains for the given regions (black rectangles in **a–c**). Grid cells over marine areas are excluded. The mean NO$_2$ and population density (pd) are provided for each domain. Population data are from the Gridded Population of the World, Version 4 (GPWv4, adjusted to Match 2015 Revision of UN WPP Country Totals) provided by the NASA Socioeconomic Data and Applications Center (SEDAC) (https://doi.org/10.7927/H4PN93PB).

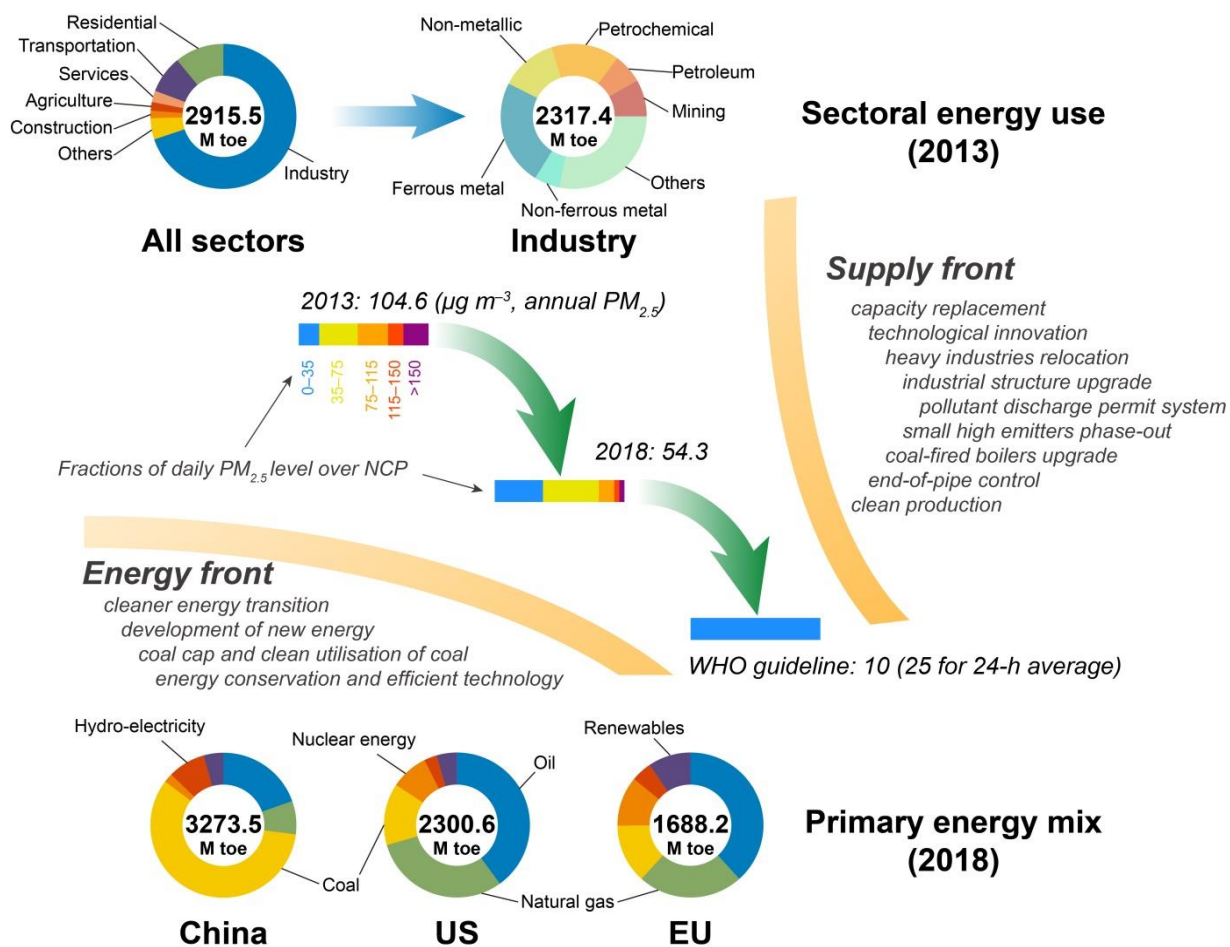

**Figure 4 | Past successes and future choices for air pollution controls in China.** Horizontal bars show the big step in the improvement of annual $PM_{2.5}$ levels over the NCP (Fig. 1a) during 2013–2018, colored by the fractions of daily $PM_{2.5}$ level classified by ambient air quality index (AQI, shown for the first five levels)[33], and the air quality goals to achieve, which requires effective policy choices on both supply and energy fronts. The top two doughnut charts show the sectoral energy use for China in 2013 (data from the National Bureau of Statistics of China, http://data.stats.gov.cn/index.htm); and the bottom three the primary energy mix in 2018 for China, the US, and the European Union (EU), respectively[35]. Numbers in doughnut charts provide the annual energy consumption amounts in million tons of oil equivalent (M toe).

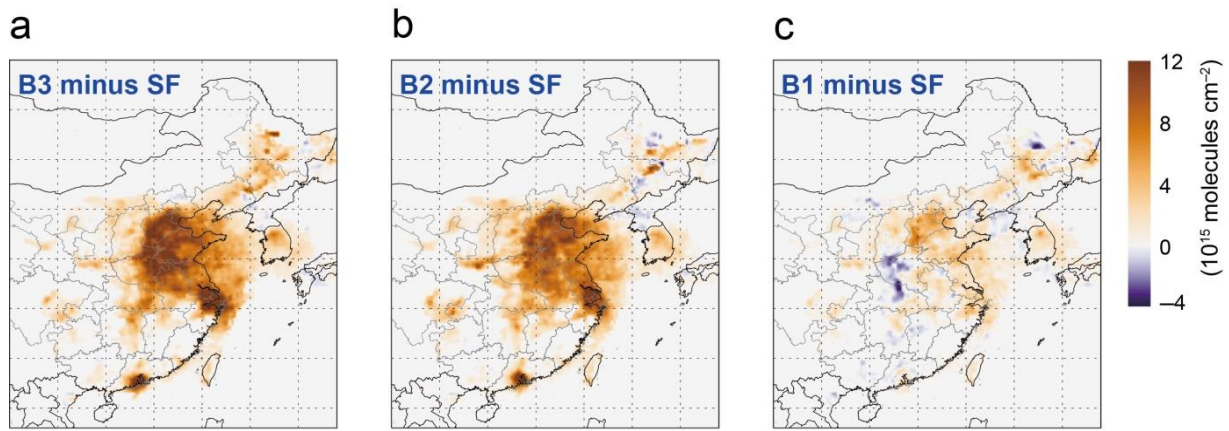

**Extended Data Figure 1 | Weekly evolution of Aura-OMI tropospheric NO₂ columns over eastern China before the SF. a–c,** Maps show the column averages of the three weeks before the SF (B3–B1) minus that of the SF week for 2005–2019 (Fig. 3a), respectively. Data are aligned and averaged according to the Chinese lunar calendar dates (Supplementary Information). Remote areas with grid average columns $<2 \times 10^{15}$ molecules cm$^{-2}$ are excluded.

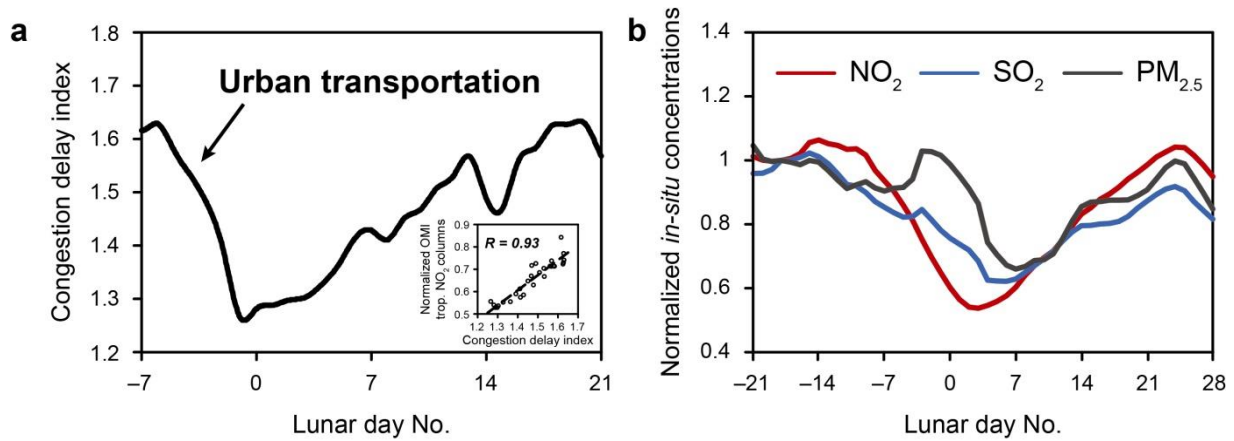

**Extended Data Figure 2 | Urban congestion delay index and *in-situ* measurements of PM₂.₅ and its precursors from MEE monitoring sites during the SF. a,** The congestion delay index of urban transportation based on the traffic data for Chinese cities[28]. The small panel shows the correlation between the normalized OMI tropospheric $NO_2$ columns (Fig. 2a) and the congestion delay index for days –7 to 21, with the linear regression fit (dashed line). **b,** Average profiles of the *in-situ* $PM_{2.5}$, $NO_2$, and $SO_2$ concentrations during 2013–2019 over the "GDP-Top-50" cities (Extended Data Table 1). Data are normalized for each city and 7-day moving-smoothed for the average profiles with baseline variations (modeling) adjustments.

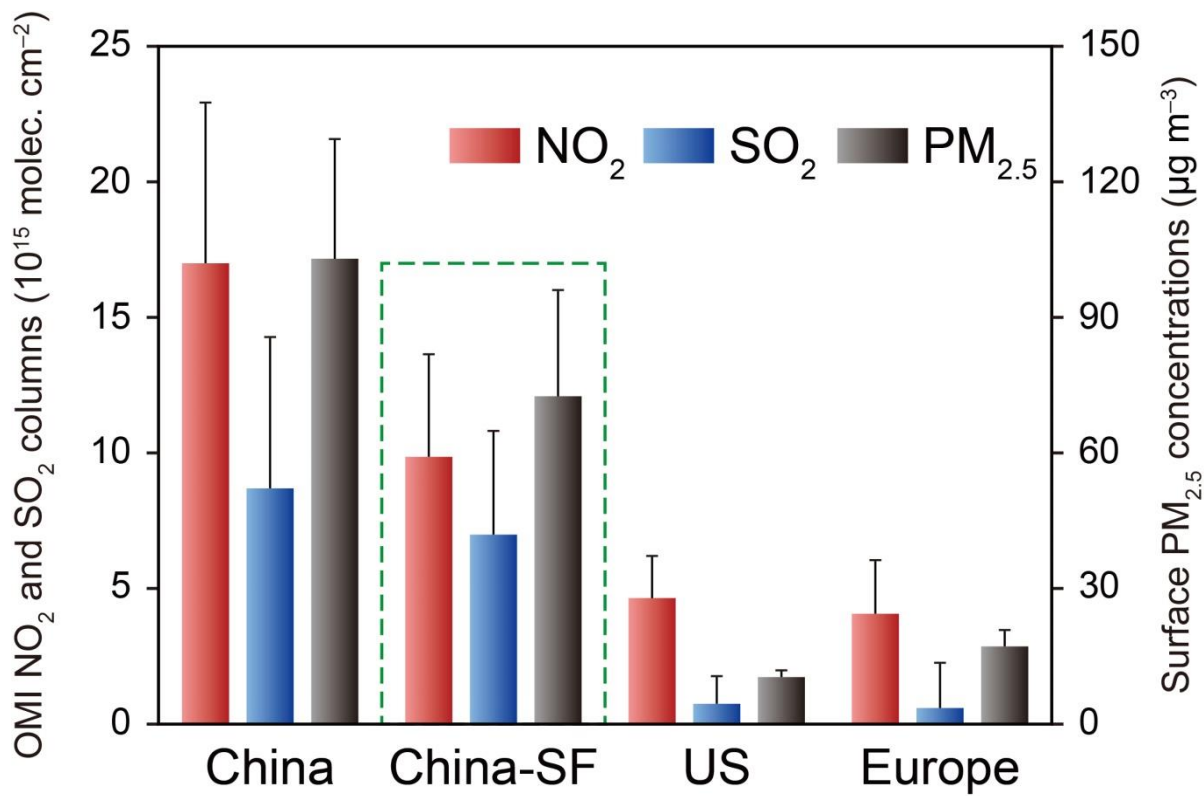

**Extended Data Figure 3 | Cross-regional comparison of air pollutant concentrations.** The OMI tropospheric $NO_2$ and $SO_2$ columns and surface $PM_{2.5}$ concentrations of the January–February averages over domains in Fig. 3**a–c** (black rectangles) for the three regions (eastern China, eastern US, and western Europe), and of the SF-week averages over the domain for eastern China (China-SF). OMI columns during the SF weeks have been excluded from the January–February averages for China. $PM_{2.5}$ data during the lunar days 7 to 13 are excluded for China to take a 7-day lag period into account. Data are averaged for 2013–2019 (European $PM_{2.5}$ for 2013–2016). The lines above the bars show the +1 standard deviation of the variables. Air quality data for the US and Europe are from the US Environmental Protection Agency (https://aqs.epa.gov/aqsweb/airdata/download_files.html) and the European Environment Agency (http://discomap.eea.europa.eu/map/fme/AirQualityExport.htm).

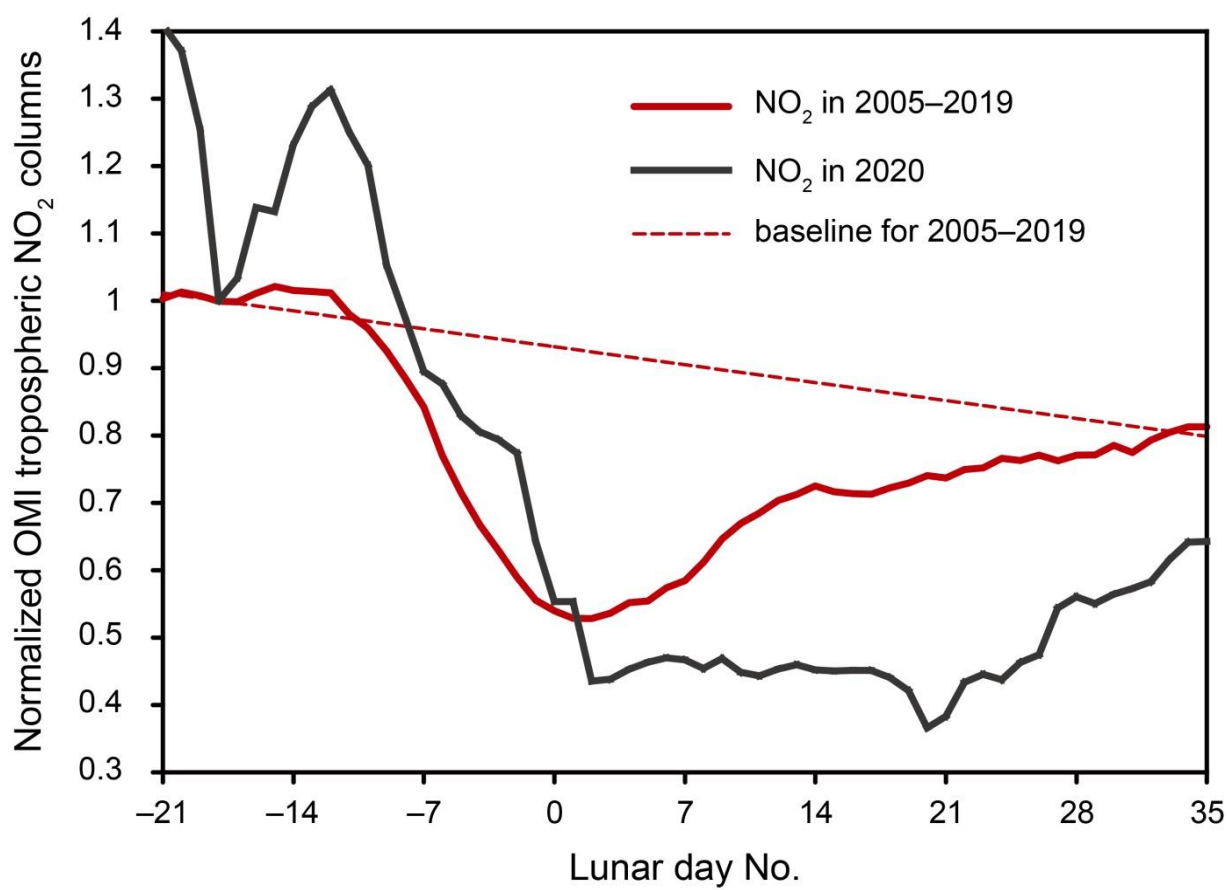

**Extended Data Figure 4 | Evolution of tropospheric NO₂ columns during the SF in 2020 over the "GDP-Top-50" cities coinciding with the outbreak of COVID-19.** The red and black solid lines show the normalized tropospheric NO$_2$ column profiles for 2005–2019 (red) and 2020 (black), respectively. The dashed baseline was generated by the chemical transport model (CTM) for 2005–2019 with fixed emission rates of air pollutants (Methods). A significant delay in NO$_2$ pollution rebound can be observed in 2020. Note that the substantial fluctuation in NO$_2$ profile of 2020 that indicates significant meteorology-induced biases when examining human activity changes based on data of a single year. Our analyses on multi-year averaged air pollution evolutions during the SF period can act as a baseline to evaluate the broad suppression of human activities during the COVID-19 event in China.

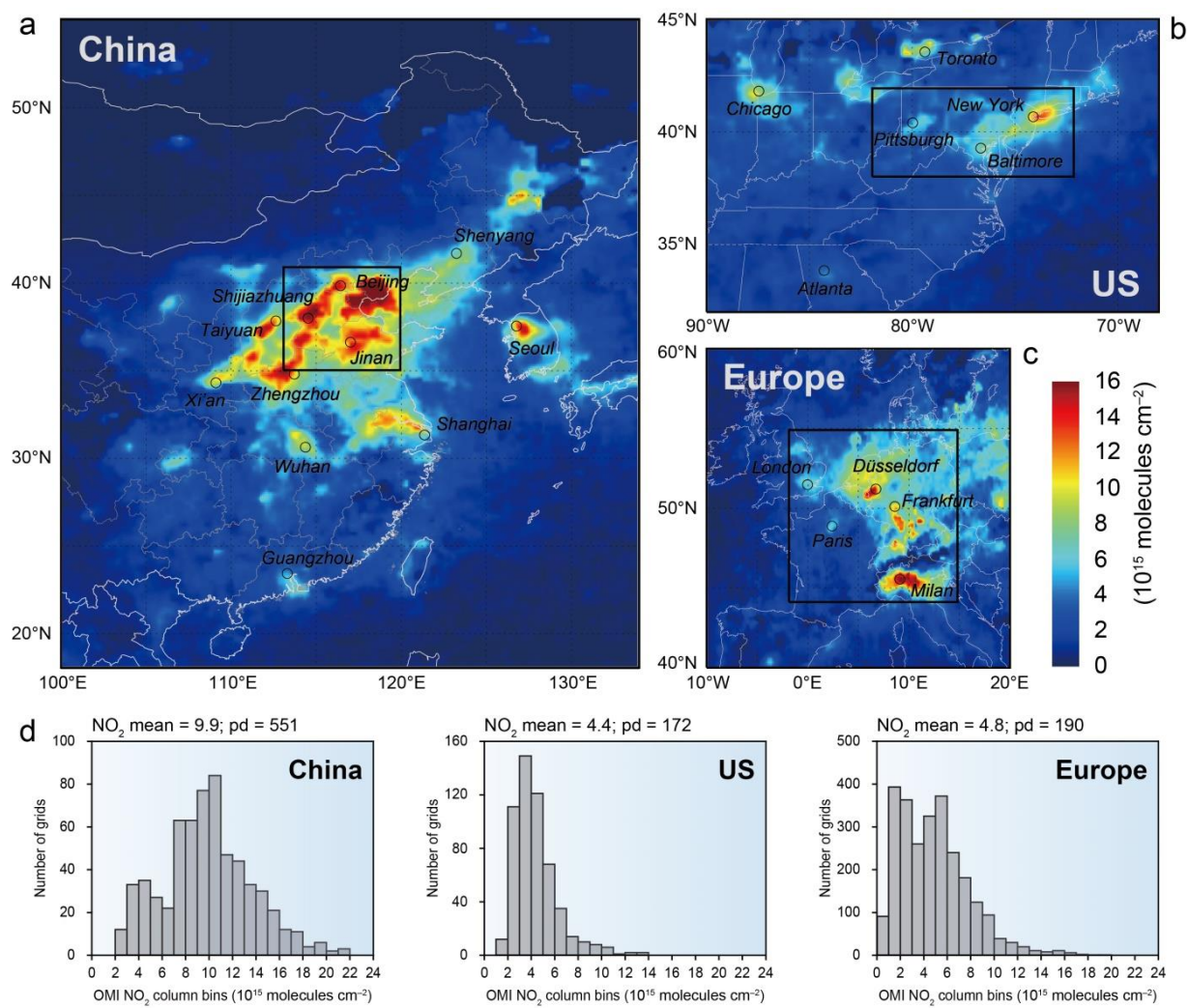

**Extended Data Figure 5 | Cross-regional comparison of NO₂ pollution for the SF weeks of 2013–2019. a–c,** Aura-OMI tropospheric NO₂ columns over three mid-latitude regions in the North Hemisphere (**a**, eastern China; **b**, eastern US; **c**, western Europe). **d,** Distribution histograms of the gridded OMI tropospheric NO₂ columns in the selected domains for the given regions (black rectangles in **a**–**c**). Data processing and symbols similar to Fig. 3.

**Extended Data Table 1 | The "GDP-Top-50" cities in China\***

| Rank | Cities | Latitudes (°N) | Longitudes (°E) | Population (million) | GDP (billion CNY) | NO$_2$ reductions (%) |
|------|--------|----------------|-----------------|----------------------|-------------------|-----------------------|
| 1 | Shanghai | 31.23 | 121.47 | 24.18 | 3013.3 | 52.7 |
| 2 | Beijing | 39.90 | 116.41 | 21.71 | 2800.0 | 31.5 |
| 3 | Shenzhen | 22.54 | 114.06 | 10.90 | 2228.6 | 49.1 |
| 4 | Guangzhou | 23.13 | 113.26 | 14.04 | 2150.0 | 46.9 |
| 5 | Chongqing | 29.43 | 106.91 | 33.72 | 1953.0 | 19.3 |
| 6 | Tianjin | 39.34 | 117.36 | 15.47 | 1859.5 | 43.4 |
| 7 | Suzhou | 31.30 | 120.59 | 10.65 | 1700.0 | 51.2 |
| 8 | Chengdu | 30.57 | 104.07 | 15.92 | 1389.0 | 27.3 |
| 9 | Wuhan | 30.59 | 114.31 | 10.77 | 1340.0 | 39.6 |
| 10 | Hangzhou | 30.27 | 120.16 | 9.19 | 1255.6 | 70.2 |
| 11 | Nanjing | 32.06 | 118.80 | 8.27 | 1171.5 | 33.6 |
| 12 | Qingdao | 36.07 | 120.38 | 8.71 | 1125.8 | 38.0 |
| 13 | Wuxi | 31.49 | 120.31 | 6.53 | 1051.1 | 49.9 |
| 14 | Changsha | 28.23 | 112.94 | 7.65 | 1020.0 | 53.4 |
| 15 | Ningbo | 29.87 | 121.54 | 7.88 | 985.0 | 59.0 |
| 16 | Foshan | 23.02 | 113.12 | 8.46 | 950.0 | 47.1 |
| 17 | Zhengzhou | 34.75 | 113.63 | 10.01 | 900.3 | 26.6 |
| 18 | Nantong | 31.98 | 120.89 | 7.30 | 775.0 | 45.6 |
| 19 | Dongguan | 23.02 | 113.75 | 8.32 | 758.0 | 59.5 |
| 20 | Yantai | 37.46 | 121.45 | 7.01 | 755.0 | 42.7 |
| 21 | Quanzhou | 24.87 | 118.68 | 8.51 | 753.3 | 21.9 |
| 22 | Dalian | 38.91 | 121.61 | 7.00 | 736.3 | 38.5 |
| 23 | Jinan | 36.65 | 117.12 | 7.06 | 728.5 | 35.6 |
| 24 | Xi'an | 34.34 | 108.94 | 9.45 | 720.6 | 32.6 |
| 25 | Hefei | 31.82 | 117.23 | 9.37 | 719.1 | 62.7 |
| 26 | Fuzhou | 26.07 | 119.30 | 7.57 | 712.8 | 45.6 |
| 27 | Tangshan | 39.63 | 118.18 | 10.24 | 701.2 | 21.4 |
| 28 | Changzhou | 31.81 | 119.97 | 4.71 | 662.0 | 44.6 |
| 29 | Changchun | 43.82 | 125.32 | 8.80 | 661.3 | 42.4 |
| 30 | Ha'erbin | 45.80 | 126.54 | 10.63 | 660.9 | 54.6 |
| 31 | Xuzhou | 34.21 | 117.28 | 8.71 | 660.0 | 50.3 |
| 32 | Shijiazhuang | 38.04 | 114.51 | 10.78 | 655.8 | 31.1 |
| 33 | Weifang | 36.71 | 119.16 | 9.27 | 632.5 | 42.8 |
| 34 | Shenyang | 41.81 | 123.43 | 8.29 | 587.0 | 19.8 |
| 35 | Wenzhou | 27.99 | 120.70 | 9.19 | 548.5 | 35.1 |
| 36 | Shaoxing | 30.00 | 120.59 | 5.01 | 531.1 | 71.2 |

Continued.

| Rank | Cities | Latitudes (°N) | Longitudes (°E) | Population (million) | GDP (billion CNY) | NO$_2$ reductions (%) |
|------|--------|---------------|-----------------|---------------------|-------------------|----------------------|
| 37 | Yangzhou | 32.39 | 119.41 | 5.00 | 506.4 | 43.8 |
| 38 | Yancheng | 33.35 | 120.16 | 7.24 | 505.0 | 45.1 |
| 39 | Nanchang | 28.68 | 115.86 | 5.37 | 500.0 | 41.7 |
| 40 | Zibo | 36.81 | 118.06 | 4.64 | 488.6 | 36.8 |
| 41 | Kunming | 24.88 | 102.83 | 7.26 | 485.6 | 16.9 |
| 42 | Taizhou | 32.46 | 119.92 | 5.08 | 474.4 | 38.7 |
| 43 | E'erduosi | 39.61 | 109.78 | 2.01 | 471.6 | 10.6 |
| 44 | Jining | 35.42 | 116.59 | 8.08 | 462.0 | 42.6 |
| 45 | Taizhou | 28.66 | 121.42 | 6.03 | 438.8 | 57.6 |
| 46 | Linyi | 35.10 | 118.36 | 10.44 | 434.5 | 38.0 |
| 47 | Luoyang | 34.62 | 112.45 | 6.80 | 434.3 | 12.6 |
| 48 | Xiamen | 24.48 | 118.09 | 4.00 | 430.0 | 46.1 |
| 49 | Dongying | 37.43 | 118.67 | 2.09 | 419.8 | 33.2 |
| 50 | Nanning | 22.82 | 108.37 | 7.52 | 418.0 | 30.6 |
| Summary | - | - | - | 33.4% | 56.0% | 40.6±13.7 |

*The gross domestic product (GDP) statistics data in 2017 are used for the ranking of Chinese cities. The reduction ratios (%) of Aura-OMI tropospheric NO$_2$ columns for the SF weeks during 2005–2019 are given (last column) for each city with consideration of the average adjustment factor (Methods). The summary provides the shares of the population and GDP of the 50 cities in the annual totals of China (%), and the average reduction ratio of tropospheric NO$_2$ columns among the 50 cities with a standard deviation.

**Extended Data Table 2 | The isolated coal-fired power plants analyzed in this study***

| Plant # | Locations | Latitudes ($^\circ$N) | Longitudes ($^\circ$E) | Capacities (MW) | NO$_2$ ratios |
|---|---|---|---|---|---|
| 1 | Wuhai | 39.32 | 106.86 | 1060 | 0.96 |
| 2 | Chifeng | 42.30 | 119.32 | 2100 | 0.91 |
| 3 | Kaiyuan | 23.81 | 103.18 | 600 | 0.96 |
| 4 | Luanhe | 40.94 | 117.76 | 1010 | 1.02 |
| 5 | Shangdu | 42.22 | 116.03 | 2400 | 0.95 |
| Summary | | - | - | - | 0.96 ±0.04 |

*The five isolated coal-fired power plants are adopted from the Grade-A plants in ref.[25], with scarce pollution interference from nearby areas. The ratios of Aura-OMI tropospheric NO$_2$ columns for the SF weeks during 2005–2019 versus the local column averages of the third week before the SF are given in the last column for each isolated power plant. The summary provides the average ratio of tropospheric NO$_2$ columns with a standard deviation.

**Extended Data Table 3 | The monthly industrial and residential electricity consumption for the first quarters of 2010–2019 in China\***

| Year | Industrial electricity consumption (billion kWh) | | | Ratios in SF-month\*\* | Residential electricity consumption (billion kWh) | | | Ratios in SF-month\*\* |
|------|------|------|------|------|------|------|------|------|
| | Jan | Feb | Mar | | Jan | Feb | Mar | |
| 2010 | 258.2 | 188.8 | 257.2 | 0.81 | 44.2 | 40.1 | 43.7 | 1.01 |
| 2011 | 285.3 | 213.1 | 291.4 | 0.82 | 47.0 | 50.1 | 48.4 | 1.16 |
| 2012 | 255.4 | 267.2 | 302.2 | 0.87 | 50.6 | 61.1 | 56.5 | 0.83 |
| 2013 | 320.5 | 221.3 | 311.7 | 0.78 | 60.1 | 56.8 | 56.6 | 1.08 |
| 2014 | 310.4 | 260.3 | 320.6 | 1.02 | 59.9 | 61.8 | 63.2 | 0.91 |
| 2015 | 342.5 | 234.7 | 313.4 | 0.79 | 64.9 | 61.0 | 64.5 | 1.04 |
| 2016 | 341.6 | 237.6 | 332.9 | 0.75 | 69.6 | 70.9 | 70.3 | 1.08 |
| 2017 | 330.1 | 290.6 | 362.7 | 0.96 | 70.1 | 76.6 | 71.3 | 0.90 |
| 2018 | 411.2 | 273.3 | 353.4 | 0.79 | 81.5 | 86.8 | 89.0 | 1.13 |
| 2019 | 402.7 | 288.1 | 383.0 | 0.81 | 93.4 | 96.5 | 93.0 | 1.15 |
| Summary | - | - | - | 0.79 ±0.02 | - | - | - | 1.09 ±0.06 |

\*The term "ratio in SF-month" (Gregorian dates of the SFs available in Supplementary Information) represents the ratio of the industrial (or residential) electricity consumption (data from ref.[29]) in the month when the SF occurs (either January or February) to that in the other two months. The discrepancies in monthly day numbers have been taken into consideration. The summary gives the average values of the ratios in the SF month with their standard deviations. \*\*Due to the occurrence of SF late in January of 2012 (Jan 23), 2014 (Jan 31) and 2017 (Jan 28), data of these three years are not taken into the calculation to avoid cross-month interference.

Supplementary Information for

**Spring Festival points the way to cleaner air in China**

Siwen Wang[1], Hang Su[1]*, David G. Streets[2], Qiang Zhang[3], Zifeng Lu[2], Kebin He[4], Meinrat O. Andreae[1,5], Ulrich Pöschl[1] and Yafang Cheng[1]*

[1]Max Planck Institute for Chemistry, Mainz 55128, Germany

[2]Energy Systems Division, Argonne National Laboratory, Lemont, IL 60439, USA

[3]Ministry of Education Key Laboratory for Earth System Modeling, Department of Earth System Science, Tsinghua University, Beijing 100084, China

[4]State Key Joint Laboratory of Environment Simulation and Pollution Control, School of Environment, Tsinghua University, Beijing 100084, China

[5]Scripps Institution of Oceanography, University of California San Diego, La Jolla, CA 92093, USA

*e-mail: yafang.cheng@mpic.de; h.su@mpic.de

## Contents

**The Spring Festival (SF)**

The Spring Festival (SF), also known as the Chinese New Year, is the most important traditional festival celebrated in China, lasting from the first day of the lunar New Year to the 15th day (the Lantern Festival, LF) or longer on Chinese lunar calendar. The Chinese lunar calendar is about 20–50 days behind the Gregorian calendar. Hence, about one-third of the SFs fall in late January, and the else begin in February. Table S4 shows the Gregorian dates of the SFs in 2005–2020.

There are 1-week national statutory holidays for the SF in China, starting from the lunar New Year's Day. Most of the Chinese return to their hometowns from ordinary living/working places for the family reunion before the holidays. It hence forms the world's largest year after year round-trip long-distance human migration (lasting for about 40 days). In 2019, the domestic transportation capacity for the SF travel is about 3 billion as reported[56]. The migration booms in the preceding week of the SF holidays and falls to the lowest on the lunar New Year's Day, and rebounds when people return. The direction of the migration is usually from the developed metropolises (e.g., the national central cities) to moderate and small cities, to towns, and to rural areas before the holidays, and on the contrary after the holidays. The real-time trajectory of the past 8-h migration flow during the SF travel can be snapshot on several real-time digital transportation data platforms based on the Location Based Services (LBS) technology[57,58].

In this study, we defined a set of lunar day No. according to the Chinese lunar calendar for days before and after the SF (totally extended to 90 days) as partly indicated in Supplementary Table 1 (the entire period using in analysis covering days –42 to 47). The lunar New Year's Day is set to day 0 and accordingly the LF is day 14. Days before the lunar New Year's Day are set to negative numbers. The SF holiday week (days 0–6) is denoted as the SF Week.

**Future applications on source structure diagnostics**

The specific $NO_2$ profiles for individual cities (or regions) during the SF are useful for inferring the emission source structure when given the activity profile or fingerprint of each major source. To a first approximation the shares of these sources in the total $NO_x$ emissions can be regarded as a set of weighting factors multiplied with the corresponding source activity profiles, to generate the integrated observed $NO_2$ profiles. This practice avoids the complex bottom-up estimation based on activity data and emission factors, which usually have large uncertainties[59]. To achieve this, it would be necessary to collect more detailed activity profiles during the SF for a range of source categories.

**Supplementary Table 1 | The Gregorian dates of the Spring Festivals in 2005–2020\***

| Year | Lunar day No. | | | | | | | | | | | | | | | | | | | | | | |
|---|---|---|---|---|---|---|---|---|---|---|---|---|---|---|---|---|---|---|---|---|---|---|---|
| | -7 | -6 | -5 | -4 | -3 | -2 | -1 | 0 | 1 | 2 | 3 | 4 | 5 | 6 | 7 | 8 | 9 | 10 | 11 | 12 | 13 | 14 | 15 |
| 2005 | 0202 | 0203 | 0204 | 0205 | 0206 | 0207 | 0208 | 0209 | 0210 | 0211 | 0212 | 0213 | 0214 | 0215 | 0216 | 0217 | 0218 | 0219 | 0220 | 0221 | 0222 | 0223 | 0224 |
| 2006 | 0122 | 0123 | 0124 | 0125 | 0126 | 0127 | 0128 | 0129 | 0130 | 0131 | 0201 | 0202 | 0203 | 0204 | 0205 | 0206 | 0207 | 0208 | 0209 | 0210 | 0211 | 0212 | 0213 |
| 2007 | 0211 | 0212 | 0213 | 0214 | 0215 | 0216 | 0217 | 0218 | 0219 | 0220 | 0221 | 0222 | 0223 | 0224 | 0225 | 0226 | 0227 | 0228 | 0301 | 0302 | 0303 | 0304 | 0305 |
| 2008 | 0131 | 0201 | 0202 | 0203 | 0204 | 0205 | 0206 | 0207 | 0208 | 0209 | 0210 | 0211 | 0212 | 0213 | 0214 | 0215 | 0216 | 0217 | 0218 | 0219 | 0220 | 0221 | 0222 |
| 2009 | 0119 | 0120 | 0121 | 0122 | 0123 | 0124 | 0125 | 0126 | 0127 | 0128 | 0129 | 0130 | 0131 | 0201 | 0202 | 0203 | 0204 | 0205 | 0206 | 0207 | 0208 | 0209 | 0210 |
| 2010 | 0207 | 0208 | 0209 | 0210 | 0211 | 0212 | 0213 | 0214 | 0215 | 0216 | 0217 | 0218 | 0219 | 0220 | 0221 | 0222 | 0223 | 0224 | 0225 | 0226 | 0227 | 0228 | 0301 |
| 2011 | 0127 | 0128 | 0129 | 0130 | 0131 | 0201 | 0202 | 0203 | 0204 | 0205 | 0206 | 0207 | 0208 | 0209 | 0210 | 0211 | 0212 | 0213 | 0214 | 0215 | 0216 | 0217 | 0218 |
| 2012 | 0116 | 0117 | 0118 | 0119 | 0120 | 0121 | 0122 | 0123 | 0124 | 0125 | 0126 | 0127 | 0128 | 0129 | 0130 | 0131 | 0201 | 0202 | 0203 | 0204 | 0205 | 0206 | 0207 |
| 2013 | 0203 | 0204 | 0205 | 0206 | 0207 | 0208 | 0209 | 0210 | 0211 | 0212 | 0213 | 0214 | 0215 | 0216 | 0217 | 0218 | 0219 | 0220 | 0221 | 0222 | 0223 | 0224 | 0225 |
| 2014 | 0124 | 0125 | 0126 | 0127 | 0128 | 0129 | 0130 | 0131 | 0201 | 0202 | 0203 | 0204 | 0205 | 0206 | 0207 | 0208 | 0209 | 0210 | 0211 | 0212 | 0213 | 0214 | 0215 |
| 2015 | 0212 | 0213 | 0214 | 0215 | 0216 | 0217 | 0218 | 0219 | 0220 | 0221 | 0222 | 0223 | 0224 | 0225 | 0226 | 0227 | 0228 | 0301 | 0302 | 0303 | 0304 | 0305 | 0306 |
| 2016 | 0201 | 0202 | 0203 | 0204 | 0205 | 0206 | 0207 | 0208 | 0209 | 0210 | 0211 | 0212 | 0213 | 0214 | 0215 | 0216 | 0217 | 0218 | 0219 | 0220 | 0221 | 0222 | 0223 |
| 2017 | 0121 | 0122 | 0123 | 0124 | 0125 | 0126 | 0127 | 0128 | 0129 | 0130 | 0131 | 0201 | 0202 | 0203 | 0204 | 0205 | 0206 | 0207 | 0208 | 0209 | 0210 | 0211 | 0212 |
| 2018 | 0209 | 0210 | 0211 | 0212 | 0213 | 0214 | 0215 | 0216 | 0217 | 0218 | 0219 | 0220 | 0221 | 0222 | 0223 | 0224 | 0225 | 0226 | 0227 | 0228 | 0301 | 0302 | 0303 |
| 2019 | 0129 | 0130 | 0131 | 0201 | 0202 | 0203 | 0204 | 0205 | 0206 | 0207 | 0208 | 0209 | 0210 | 0211 | 0212 | 0213 | 0214 | 0215 | 0216 | 0217 | 0218 | 0219 | 0220 |
| 2020 | 0118 | 0119 | 0120 | 0121 | 0122 | 0123 | 0124 | 0125 | 0126 | 0127 | 0128 | 0129 | 0130 | 0131 | 0201 | 0202 | 0203 | 0204 | 0205 | 0206 | 0207 | 0208 | 0209 |

\*The format of the Gregorian date is mmdd. Based on the Chinese lunar calendar, the lunar New Year's is set to day 0 in this study, and accordingly the Lantern Festival is day 14.